\documentclass[aps,prb,twocolumn,a4paper,superscriptaddress,longbibliography]{revtex4-2}

\usepackage{graphicx}
\usepackage{amssymb}
\usepackage{hyperref}

\begin{document}

\title{Angular dependence of the low-temperature skyrmion state in Cu$_2$OSeO$_3$}

\author{Christian Oberleitner}
\affiliation{Physics Department, Technical University of Munich, 85748 Garching, Germany}

\author{Andreas Bauer}
\email{andreas.bauer@ph.tum.de}
\affiliation{Physics Department, Technical University of Munich, 85748 Garching, Germany}
\affiliation{Center for QuantumEngineering (ZQE), Technical University of Munich, 85748 Garching, Germany}

\author{Lukas Heindl}
\affiliation{Physics Department, Technical University of Munich, 85748 Garching, Germany}

\author{Lin Chen}
\affiliation{Physics Department, Technical University of Munich, 85748 Garching, Germany}

\author{Aisha Aqeel}
\affiliation{Experimental Physics IV, University of Augsburg, 86159 Augsburg, Germany}
\affiliation{Munich Center for Quantum Science and Technology (MCQST), 80799 Munich, Germany}

\author{Christian H. Back}
\affiliation{Physics Department, Technical University of Munich, 85748 Garching, Germany}
\affiliation{Center for QuantumEngineering (ZQE), Technical University of Munich, 85748 Garching, Germany}
\affiliation{Munich Center for Quantum Science and Technology (MCQST), 80799 Munich, Germany}

\author{Hans Huebl}
\affiliation{Physics Department, Technical University of Munich, 85748 Garching, Germany}
\affiliation{Munich Center for Quantum Science and Technology (MCQST), 80799 Munich, Germany}
\affiliation{Walther-Mei{\ss}ner-Institut, Bayerische Akademie der Wissenschaften, 85748 Garching, Germany}

\author{Christian Pfleiderer}
\affiliation{Physics Department, Technical University of Munich, 85748 Garching, Germany}
\affiliation{Center for QuantumEngineering (ZQE), Technical University of Munich, 85748 Garching, Germany}
\affiliation{Munich Center for Quantum Science and Technology (MCQST), 80799 Munich, Germany}
\affiliation{Heinz Maier-Leibnitz Zentrum (MLZ), Technical University of Munich, 85748 Garching, Germany}

\date{\today}

\begin{abstract}
In the cubic chiral magnet Cu$_2$OSeO$_3$, a low-temperature skyrmion state (LTS) and a metastable tilted conical state emerge for magnetic field along $\langle 100\rangle$, both stabilized by magnetocrystalline anisotropies. We report how these states evolve when the magnetic field is rotated from $\langle 100\rangle$ via $\langle 111\rangle$ to $\langle 110\rangle$. Measurements of the magnetization and the AC susceptibility are combined with broadband microwave spectroscopy on the same cuboid sample at 5\,K. In the microwave spectra, the characteristic gigahertz excitations, notably the counterclockwise gyration mode of the skyrmions, provide a direct indicator of the character of the magnetic states and a qualitative measure of their volume fractions. Both techniques consistently show that signatures of the LTS decrease in strength under rotation and are observed only for field directions within ${\sim}20^{\circ}$ of $\langle 100\rangle$. In this range, the field values at which the skyrmions unwind remain essentially unchanged. In contrast, signatures of the tilted conical state persist up to ${\sim}40^{\circ}$, beyond which the conventional sequence of helical, conical, and field-polarized states is recovered. Thus, the tilted conical state, while being a prerequisite for the nucleation of the LTS under field sweeps, is not sufficient for its formation.
\end{abstract}

\maketitle

\section{Introduction}

In cubic chiral magnets crystallizing in space group $P2_13$, such as MnSi, Fe$_{1-x}$Co$_x$Si, FeGe, and Cu$_2$OSeO$_3$, skyrmion lattices form in a small pocket of the magnetic phase diagram just below the onset of helimagnetic order~\cite{2009_Muhlbauer_Science, 2010_Munzer_PhysRevB, 2010_Yu_Nature, 2011_Yu_NatMater, 2012_Seki_Science, 2012_Adams_PhysRevLett, 2013_Nagaosa_NatNanotechnol}. For the stabilization of this high-temperature skyrmion lattice, thermal fluctuations are key~\cite{2009_Muhlbauer_Science, 2013_Buhrandt_PhysRevB}, and the field orientation only subtly influences the orientation of the skyrmion lattice, owing to weak magnetocrystalline anisotropies~\cite{2012_Adams_PhysRevLett, 2012_Seki_PhysRevB, 2017_Makino_PhysRevB, 2017_Bannenberg_PhysRevB, 2018_Adams_PhysRevLett, 2016_Bauer_Book}.

In Cu$_2$OSeO$_3$, additionally a skyrmion state emerges at low temperatures close to the transition between the conical and the field-polarized state, accompanied by a tilted conical state in which the propagation directions of conical domains tilt away from the field direction~\cite{2018_Chacon_NatPhys, 2018_Qian_SciAdv, 2018_Halder_PhysRevB, 2019_Bannenberg_npjQuantumMater, 2023_Crisanti_PhysRevRes, 2024_Marchiori_CommunMater_202, 2025_Mehboodi_SciTechnolAdvMater}. This low-temperature skyrmion state (LTS) and the tilted conical state are observed for field along the easy $\langle 100\rangle$ axes, but not along $\langle 110\rangle$ or $\langle 111\rangle$~\cite{2019_Prasai_PhysRevB, 2020_Leonov_PhysRevRes}, indicating that magnetocrystalline anisotropies are crucial, possibly assisted by exchange anisotropies~\cite{2020_Birch_PhysRevB, 2021_Moody_PhysRevRes, 2023_Baral_PhysRevRes, 2023_Leonov_FrontPhys}. Although the LTS represents the ground state in a finite field range, its nucleation at low temperatures is kinetically limited, resulting in a pronounced dependence on the temperature and field history~\cite{2018_Chacon_NatPhys, 2021_Aqeel_PhysRevLett}. The tilted conical state always forms before the LTS and is considered an intermediate state for its nucleation, potentially connected with the presence of domain boundaries~\cite{2018_Halder_PhysRevB, 2022_Leonov_PhysRevRes, 2024_Marchiori_CommunMater_202}. Accordingly, cycling the field within the tilted conical state increases the volume fraction of the LTS~\cite{2018_Chacon_NatPhys, 2021_Aqeel_PhysRevLett, 2022_Lee_JPhysCondensMatter}.

Given the decisive role of the anisotropies, the dependence of the tilted conical state and the LTS on the field direction is an obvious question. So far, however, the magnetic phase diagram has been determined only for the high-symmetry directions, and the LTS has not been studied for oblique fields. In small-angle neutron scattering on Zn-substituted Cu$_2$OSeO$_3$, rotating the field at 5\,K from $\langle 110\rangle$ via $\langle 111\rangle$ to $\langle 100\rangle$ revealed increasing deviations of the conical propagation direction from the field direction close to $\langle 100\rangle$, attributed to the tilted conical state, while the LTS was not addressed~\cite{2021_Moody_PhysRevRes}. Further, an LTS nucleated for field along $\langle 100\rangle$ was found to persist when the crystal was subsequently rotated by up to $40^{\circ}$~\cite{2018_Chacon_NatPhys}. The range of field directions in which the tilted conical state and the LTS form, however, has not been determined.

Here, we address this question at 5\,K for field directions from $\langle 100\rangle$ via $\langle 111\rangle$ to $\langle 110\rangle$, combining magnetometry and broadband microwave spectroscopy on the same sample. The microwave spectra complement the phase boundaries inferred from magnetometry by a direct fingerprint of the magnetic states in terms of their characteristic gigahertz excitations~\cite{2012_Mochizuki_PhysRevLett, 2012_Onose_PhysRevLett, 2015_Schwarze_NatMater}. In particular, the counterclockwise gyration mode of the skyrmions signals the presence of the LTS and provides a qualitative measure of its volume fraction~\cite{2021_Aqeel_PhysRevLett, 2022_Lee_JPhysCondensMatter}. Both techniques consistently show that signatures of the LTS are confined to field directions within ${\sim}20^{\circ}$ of $\langle 100\rangle$, whereas the tilted conical state persists up to ${\sim}40^{\circ}$.


\section{Experimental methods}

A single crystal of Cu$_2$OSeO$_3$ was grown by chemical vapor transport~\cite{2010_Belesi_PhysRevB, 2010_Gnezdilov_LowTempPhys}. A cuboid with dimensions $1.45 \times 1.45 \times 1.25\,\mathrm{mm}^3$ was cut, oriented by means of x-ray Laue diffraction, and polished. The edges are parallel to [001] (1.45\,mm), $[1\bar{1}0]$ (1.45\,mm), and [110] (1.25\,mm). In both experiments, the direction of the magnetic field with respect to the crystal was varied within the $(1\bar{1}0)$ plane, i.e., by a rotation about $[1\bar{1}0]$, from [001] ($\alpha = 0^{\circ}$) to [110] ($\alpha = 90^{\circ}$), passing through [111] at $\alpha \approx 54.7^{\circ}$.

\subsection{Magnetization and AC susceptibility}

\begin{figure*}[!tbp]
	\includegraphics[width=\textwidth]{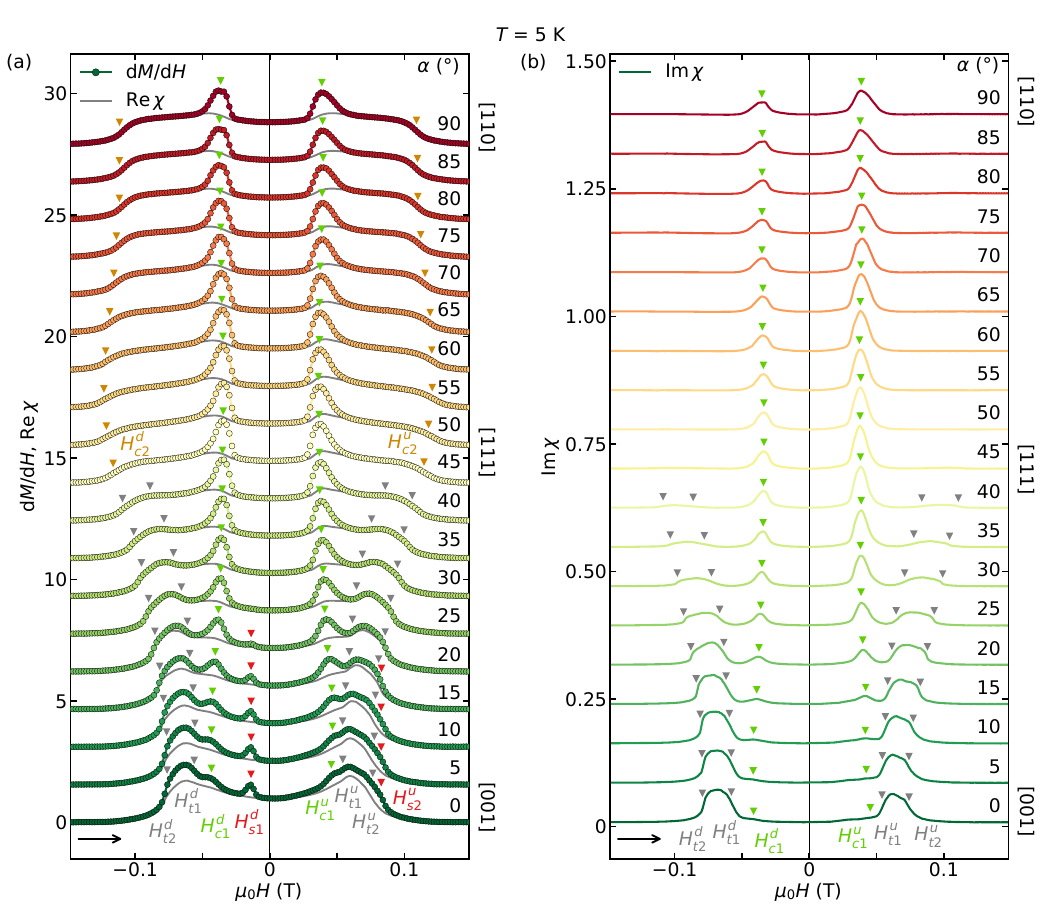}
	\caption{\label{fig:waterfall}Susceptibility of Cu$_2$OSeO$_3$ at 5\,K as a function of magnetic field for different field directions, where $\alpha$ denotes the angle between the field and $\langle 100\rangle$. Black horizontal arrows indicate the sweep direction. (a)~Differential susceptibility, $\mathrm{d}M/\mathrm{d}H$ (colored symbols), and real part of the AC susceptibility, $\mathrm{Re}\,\chi$ (gray lines). Data are offset for clarity. Colored triangles mark characteristic fields as described in the text. (b)~Imaginary part of the AC susceptibility, $\mathrm{Im}\,\chi$.}
\end{figure*}

The magnetization and the AC susceptibility were measured in a Quantum Design MPMS3 SQUID magnetometer equipped with a rotator. The magnetization was determined with an extraction technique and the differential susceptibility, $\mathrm{d}M/\mathrm{d}H$, was calculated numerically from the magnetization data. The AC susceptibility was measured at an excitation frequency of 120\,Hz and an excitation amplitude of 0.5\,mT, using 3-point scans with an integration time of 2\,s. Both the static and the excitation field were parallel to each other for all angles $\alpha$, which were varied from $-40^{\circ}$ to $95^{\circ}$ in steps of $5^{\circ}$. For clarity, Fig.~\ref{fig:waterfall} shows data for $0^{\circ} \leq \alpha \leq 90^{\circ}$ only. Both the real and the imaginary part of the AC susceptibility, $\mathrm{Re}\,\chi$ and $\mathrm{Im}\,\chi$, were considered. All data are shown as a function of the applied magnetic field, i.e., without correction for demagnetization effects.

For each angle, the sample was heated to 70\,K, i.e., well above the helimagnetic transition temperature $T_c \approx 58$\,K, and cooled to 5\,K in zero magnetic field. Data were recorded first under increasing field from 0 to $+150$\,mT and subsequently in a sweep from $+150$\,mT to $-150$\,mT. The field was changed in steps of 1\,mT and kept constant during each measurement. As the second sweep starts in the field-polarized state, its temperature and field history corresponds to high-field cooling in the terminology of Ref.~\cite{2018_Halder_PhysRevB}. For direct comparison with the microwave spectra, which were recorded from negative to positive fields, the data of the second sweep are presented with inverted field and magnetization, i.e., as a sweep from $-150$\,mT to $+150$\,mT. Following Ref.~\cite{2018_Halder_PhysRevB}, characteristic fields observed under decreasing and increasing absolute field strength (negative and positive fields in the figures) are denoted by the superscripts ``d'' and ``u'', respectively. Data recorded during the initial sweep after zero-field cooling reproduce the signatures observed under increasing absolute field strength (not shown).

\subsection{Microwave spectroscopy}

Broadband microwave spectroscopy was carried out on the same sample at 5\,K using a variable temperature insert in a two-axis vector magnet. The sample was placed on a coplanar waveguide (CPW) with a signal line of 1\,mm width, providing a predominantly in-plane excitation field. The sample was placed with its [001] edge parallel to the signal line and its short [110] edge parallel to the normal of the CPW plane. The static magnetic field was rotated by means of the vector magnet within the plane spanned by the signal line and the CPW normal, i.e., from parallel to the signal line ($\alpha = 0^{\circ}$) to perpendicular to the CPW plane ($\alpha = 90^{\circ}$). Consequently, the in-plane component of the excitation field, which is perpendicular to the signal line and parallel to the rotation axis $[1\bar{1}0]$, is perpendicular to the static field for all angles $\alpha$. As the lateral dimensions of the sample and the CPW are comparable, a weak out-of-plane component of the excitation field is expected, which is perpendicular to the static field for $\alpha = 0^{\circ}$ and parallel to it for $\alpha = 90^{\circ}$.

The complex transmission parameter $S_{21}$ was measured with a vector network analyzer at frequencies from 1 to 15\,GHz in steps of 7\,MHz. Background contributions were removed by normalizing to a reference spectrum, $S_{21}^{\mathrm{ref}}$, recorded at 700\,mT, where no resonances occur in the frequency range of interest. Spectra were recorded while sweeping the magnetic field from $-220$\,mT to $+220$\,mT in steps of 2\,mT, i.e., starting in the field-polarized state, for angles from $0^{\circ}$ to $90^{\circ}$ in steps of $5^{\circ}$.

For angles $\alpha \leq 30^{\circ}$, a second series of field sweeps was carried out in which the field was cycled in the tilted conical state to promote the nucleation of the LTS~\cite{2018_Chacon_NatPhys, 2021_Aqeel_PhysRevLett, 2022_Lee_JPhysCondensMatter}. For each angle, a cycling field $H_{\mathrm{cyc}}$ was chosen in the tilted conical state close to its high-field boundary. Upon reaching $H_{\mathrm{cyc}}$ under decreasing absolute field strength, the field was cycled 100 times between $H_{\mathrm{cyc}} - 4$\,mT and $H_{\mathrm{cyc}} + 4$\,mT at a rate of 100\,mT/min, before the field sweep was continued.

\section{Results}

\subsection{Magnetization and AC susceptibility}

Figure~\ref{fig:waterfall}(a) shows the differential susceptibility, $\mathrm{d}M/\mathrm{d}H$ (colored symbols), and the real part of the AC susceptibility, $\mathrm{Re}\,\chi$ (gray lines), for field directions between $\langle 100\rangle$ ($\alpha = 0^{\circ}$) and $\langle 110\rangle$ ($\alpha = 90^{\circ}$). The corresponding imaginary part, $\mathrm{Im}\,\chi$, is shown in Fig.~\ref{fig:waterfall}(b).

For field along $\langle 100\rangle$ (bottom curves), the data are consistent with previous reports~\cite{2018_Halder_PhysRevB, 2021_Aqeel_PhysRevLett}. Starting in the field-polarized state (FP) at large negative fields, all susceptibilities are small. Under decreasing field strength, an increase of $\mathrm{d}M/\mathrm{d}H$ at $H_{t2}^{d}$ marks the transition into the tilted conical state (T). In this field range, $\mathrm{Re}\,\chi$ is distinctly smaller than $\mathrm{d}M/\mathrm{d}H$ and $\mathrm{Im}\,\chi$ exhibits a broad maximum, characteristic of slow, dissipative processes associated with the tilted conical state~\cite{2014_Levatic_PhysRevB, 2016_Qian_PhysRevB, 2018_Halder_PhysRevB}. The boundaries $H_{t2}^{d}$ and $H_{t1}^{d}$ of the tilted conical state are inferred from the points of inflection of $\mathrm{Im}\,\chi$~\cite{2018_Halder_PhysRevB}. Below $H_{t1}^{d}$, the remaining volume fraction of tilted conical state is converted into the conical state (C) propagating along the field. Under further decreasing field, a maximum at $H_{c1}^{d}$ marks the transition into the helical state (H). Such a signature was not resolved in the cuboid sample with a different aspect ratio studied in Ref.~\cite{2018_Halder_PhysRevB}, presumably reflecting the sensitivity of the domain populations to the sample shape and the distribution of demagnetization fields. The skyrmions persist down to $H_{s1}^{d}$, well below $H_{t1}^{d}$ and $H_{c1}^{d}$, where they unwind into helical domains propagating perpendicular to the field~\cite{2013_Milde_Science, 2017_Wild_SciAdv, 2018_Halder_PhysRevB}. This unwinding is accompanied by a pronounced maximum in $\mathrm{d}M/\mathrm{d}H$ that is neither tracked by $\mathrm{Re}\,\chi$ nor accompanied by a contribution in $\mathrm{Im}\,\chi$, indicating processes that are slow compared to the AC excitation~\cite{2018_Halder_PhysRevB, 2017_Wild_SciAdv}.

Under increasing field strength (positive fields), four signatures are observed in $\mathrm{d}M/\mathrm{d}H$. These are a small maximum at $H_{c1}^{u}$ marking the transition from the helical (H) to the conical state, points of inflection at $H_{t1}^{u}$ and $H_{t2}^{u}$ bounding the tilted conical state, which are also observed in $\mathrm{Im}\,\chi$, and a shoulder at $H_{s2}^{u}$ associated with the unwinding of the skyrmions into the field-polarized state~\cite{2018_Halder_PhysRevB}. As the helical, conical, and tilted conical states are located in close proximity and the transitions extend across sizable field ranges, coexistence of different magnetic textures is expected in this field range~\cite{2006_Grigoriev_PhysRevB, 2012_Bauer_PhysRevB, 2017_Bauer_PhysRevB, 2020_Birch_PhysRevB, 2024_Marchiori_CommunMater_202, 2025_Mehboodi_SciTechnolAdvMater}.

The evolution under rotation of the field is summarized as follows. First, with increasing $\alpha$, both $H_{t1}$ and $H_{t2}$ shift to larger absolute values, e.g., from $|H_{t2}^{d}| = 76$\,mT at $0^{\circ}$ to $109$\,mT at $40^{\circ}$, while the field range of the tilted conical state of about 20 to 25\,mT remains essentially unchanged. At the same time, the broad maxima in $\mathrm{d}M/\mathrm{d}H$ and $\mathrm{Im}\,\chi$ characteristic of the tilted conical state decrease in magnitude and vanish between $40^{\circ}$ and $45^{\circ}$. For larger angles, $H_{t2}$ continuously evolves into the transition field between the conical and the field-polarized state, $H_{c2}$. $|H_{c2}|$ increases further up to a maximum of $123$\,mT at $\alpha = 55^{\circ}$, i.e., close to $\langle 111\rangle$, and decreases again to $112$\,mT towards $\langle 110\rangle$, consistent with the anisotropy of $H_{c2}$ reported for the high-symmetry directions~\cite{2018_Halder_PhysRevB, 2015_Grigoriev_PhysRevB}.

Second, the helical-to-conical transition at $H_{c1}$, observed both in $\mathrm{d}M/\mathrm{d}H$ and $\mathrm{Im}\,\chi$, shifts from $44$\,mT at $0^{\circ}$ to $36$\,mT at $35^{\circ}$, i.e., to smaller absolute values as long as the tilted conical state is present, and remains essentially constant for larger angles up to $\langle 110\rangle$.

Third, the signatures associated with the unwinding of skyrmions remain at essentially the same field values under rotation, namely $|H_{s1}^{d}| = 14$\,mT and $H_{s2}^{u} = 83$\,mT, but become less pronounced with increasing $\alpha$. The shoulder at $H_{s2}^{u}$, which is weak already for $\alpha = 0^{\circ}$, is resolved up to $15^{\circ}$, while the maximum at $H_{s1}^{d}$ decreases continuously and is resolved up to $20^{\circ}$. For $\alpha \geq 25^{\circ}$, no signatures of skyrmions are observed.

\subsection{Magnetic phase diagram}

\begin{figure}[!tbp]
	\includegraphics[width=\columnwidth]{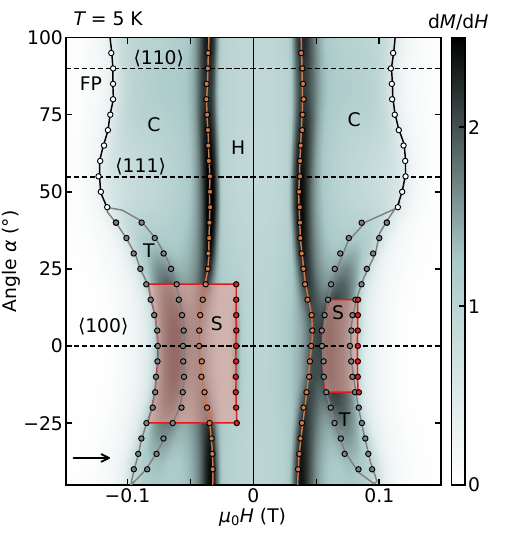}
	\caption{\label{fig:phase_dia}Magnetic phase diagram of Cu$_2$OSeO$_3$ at 5\,K as a function of the field direction, $\alpha$, as inferred from data recorded after high-field cooling. The color map depicts the differential susceptibility, $\mathrm{d}M/\mathrm{d}H$. Symbols denote characteristic fields and red shading indicates the presence of skyrmions. The following states are distinguished: field-polarized (FP), tilted conical (T), skyrmion (S), helical (H), and conical (C). The black horizontal arrow indicates the sweep direction.}
\end{figure}

The characteristic fields are summarized in the magnetic phase diagram shown in Fig.~\ref{fig:phase_dia} as a function of field direction, where the color map depicts $\mathrm{d}M/\mathrm{d}H$. As the transitions observed in $\mathrm{d}M/\mathrm{d}H$ and $\mathrm{Im}\,\chi$ coincide, only the values inferred from $\mathrm{d}M/\mathrm{d}H$ are shown. Data for negative angles, i.e., equivalent field directions, reproduce the behavior qualitatively. Quantitatively, the characteristic fields of the tilted conical state for $-\alpha$ correspond to those for $+\alpha$ at slightly smaller angles, and $H_{s1}^{d}$ is still resolved at $-25^{\circ}$ but not at $+25^{\circ}$. These asymmetries are consistent with an offset of the angular scale of about $2^{\circ}$ to $3^{\circ}$, which therefore represents the uncertainty of the angular limits stated in the following. Three angular regimes may be distinguished.

For $\alpha \lesssim 20^{\circ}$, both the tilted conical state and the LTS are observed. The LTS displays pronounced hysteresis, characteristic of its kinetically limited nucleation and topological stability~\cite{2018_Chacon_NatPhys, 2018_Halder_PhysRevB}. Under decreasing field strength, skyrmions form in the tilted conical state and persist deep into the helical state down to $H_{s1}^{d}$. Under increasing field strength, signatures of skyrmions are observed in a narrow field range between the tilted conical and the field-polarized state, terminating at $H_{s2}^{u}$.

For $25^{\circ} \lesssim \alpha \lesssim 40^{\circ}$ ($20^{\circ} \lesssim \alpha \lesssim 40^{\circ}$ under increasing field strength), signatures of the tilted conical state are still observed, while no signatures of the LTS are resolved. Here, the boundaries $H_{t1}$ and $H_{t2}$ shift to larger absolute values and the dissipative maxima in $\mathrm{Im}\,\chi$ decrease until the tilted conical state vanishes between $40^{\circ}$ and $45^{\circ}$.

For $\alpha \gtrsim 45^{\circ}$, the conventional sequence of field-polarized, conical, and helical states is observed, as known for field along $\langle 111\rangle$ and $\langle 110\rangle$~\cite{2012_Seki_Science, 2012_Adams_PhysRevLett, 2012_Seki_PhysRevB}. While the field range of the helical state is essentially independent of the field direction, the field range of the conical state is largest close to $\langle 111\rangle$, reflecting the magnetically hard $\langle 111\rangle$ axes as implied on the level of the cubic anisotropies~\cite{1980_Bak_JPhysC, 1980_Nakanishi_SolidStateCommun, 1981_Plumer_JPhysC}.

\subsection{Microwave spectroscopy for field along $\langle 100\rangle$}

\begin{figure}[!tbp]
	\includegraphics[width=\columnwidth]{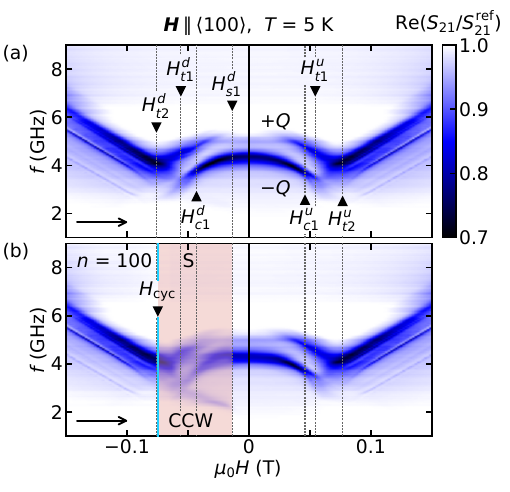}
	\caption{\label{fig:0_degree_FMR}Microwave spectra of Cu$_2$OSeO$_3$ as a function of magnetic field at 5\,K for field along $\langle 100\rangle$, starting in the field-polarized state at large negative field. The color encodes the real part of the normalized transmission. (a)~Data recorded without field cycling. (b)~Data recorded with field cycling at $H_{\mathrm{cyc}}$ (vertical turquoise line). Dotted lines indicate characteristic fields inferred from the differential susceptibility. Red shading marks the field range in which skyrmions are present. The black horizontal arrow indicates the sweep direction.}
\end{figure}

Figure~\ref{fig:0_degree_FMR}(a) shows microwave spectra in terms of the real part of the normalized transmission, $\mathrm{Re}(S_{21}/S_{21}^{\mathrm{ref}})$, for field along $\langle 100\rangle$ without field cycling. The corresponding data recorded with field cycling are shown in Fig.~\ref{fig:0_degree_FMR}(b). Dark colors correspond to resonant absorption. For comparison, characteristic fields inferred from the differential susceptibility are indicated by dotted lines.

Without field cycling, the spectra in the field-polarized state at large negative fields are dominated by the Kittel mode with its linear field dependence. Owing to the low damping of Cu$_2$OSeO$_3$ at low temperatures, on the order of $10^{-4}$~\cite{2017_Stasinopoulos_ApplPhysLett}, several additional resonances are resolved that run parallel to the Kittel mode at lower frequencies. These resonances are attributed to standing spin-wave modes~\cite{2021_Aqeel_PhysRevLett}. Under decreasing field strength, the Kittel mode exhibits a change of slope at $H_{t2}^{d}$, marking the onset of the tilted conical state~\cite{2021_Aqeel_PhysRevLett}. Below $H_{t2}^{d}$, two branches are observed, which we attribute to the $+Q$ and $-Q$ modes of the conical and tilted conical states~\cite{2015_Schwarze_NatMater, 2017_Weiler_PhysRevLett, 2021_Aqeel_PhysRevLett}. The $+Q$ mode evolves continuously from the Kittel mode, while the $-Q$ mode emerges at lower frequencies. Towards zero field, spectral weight is transferred from the $+Q$ to the $-Q$ mode, consistent with theoretical predictions for the helimagnetic excitations~\cite{2015_Schwarze_NatMater, 2017_Garst_JPhysD, 2017_Stasinopoulos_SciRep}. Under increasing field strength, the spectra evolve essentially symmetrically, i.e., without field cycling no excitations characteristic of skyrmions are resolved.

After 100 field cycles at $H_{\mathrm{cyc}}$ (turquoise line), the spectra change markedly, as shown in Fig.~\ref{fig:0_degree_FMR}(b). Spectral weight is redistributed from the $\pm Q$ modes to a distinct mode at lower frequencies. In agreement with Refs.~\cite{2021_Aqeel_PhysRevLett, 2022_Lee_JPhysCondensMatter}, we attribute this mode to the counterclockwise gyration mode of the LTS: it is located well below the $\pm Q$ modes, its frequency decreases under decreasing field strength, as also known from the high-temperature skyrmion lattice~\cite{2012_Mochizuki_PhysRevLett, 2012_Onose_PhysRevLett, 2015_Schwarze_NatMater, 2020_Seki_NatCommun}, and it vanishes at the low-field boundary of the LTS, $H_{s1}^{d}$, inferred from magnetometry. The red shading marks the corresponding field range. Under increasing field strength, where no field cycling was carried out, no excitations characteristic of skyrmions are resolved.

\subsection{Angular dependence of the microwave spectra}

\begin{figure}[!tbp]
	\includegraphics[width=\columnwidth]{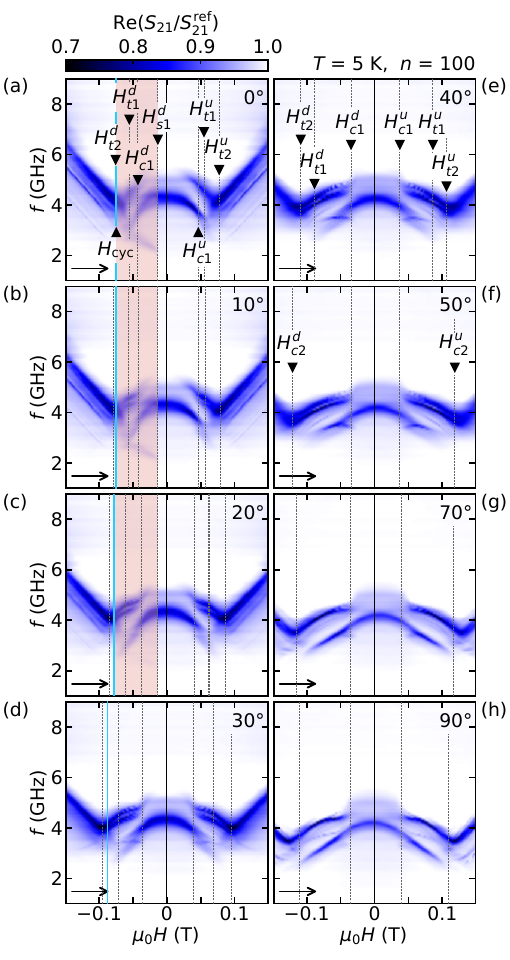}
	\caption{\label{fig:FMR_all}Microwave spectra of Cu$_2$OSeO$_3$ as a function of magnetic field at 5\,K for different field directions $\alpha$, starting in the field-polarized state at large negative field. The color encodes the real part of the normalized transmission. (a)--(d)~Data recorded with $n = 100$ field cycles at $H_{\mathrm{cyc}}$ (vertical turquoise lines) for $\alpha = 0^{\circ}$, $10^{\circ}$, $20^{\circ}$, and $30^{\circ}$. (e)--(h)~Data recorded without field cycling for $\alpha = 40^{\circ}$, $50^{\circ}$, $70^{\circ}$, and $90^{\circ}$. Dotted lines indicate characteristic fields inferred from the differential susceptibility. Red shading marks the field range in which skyrmions are present. Black horizontal arrows indicate the sweep direction.}
\end{figure}

\begin{figure}[!tbp]
	\includegraphics[width=\columnwidth]{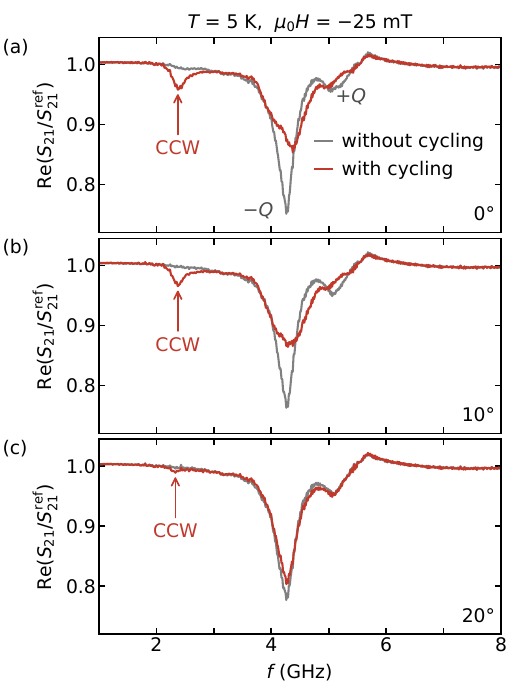}
	\caption{\label{fig:linecuts}Microwave spectra of Cu$_2$OSeO$_3$ at 5\,K and $\mu_0H = -25$\,mT under decreasing field strength, recorded without (gray) and with field cycling (red) for (a)~$\alpha = 0^{\circ}$, (b)~$10^{\circ}$, and (c)~$20^{\circ}$. Arrows mark the counterclockwise gyration mode of the skyrmions (CCW). The $-Q$ and $+Q$ modes are labeled in panel~(a).}
\end{figure}

Figure~\ref{fig:FMR_all} shows microwave spectra for $\alpha = 0^{\circ}$, $10^{\circ}$, $20^{\circ}$, and $30^{\circ}$ with field cycling [Figs.~\ref{fig:FMR_all}(a)--\ref{fig:FMR_all}(d)] and for $\alpha = 40^{\circ}$, $50^{\circ}$, $70^{\circ}$, and $90^{\circ}$ without field cycling [Figs.~\ref{fig:FMR_all}(e)--\ref{fig:FMR_all}(h)]. At $10^{\circ}$, the counterclockwise gyration mode is clearly observed after field cycling, decreasing in frequency under decreasing field strength and vanishing at $H_{s1}^{d}$, as for $0^{\circ}$. At $20^{\circ}$, the mode is weak but clearly resolved. This is illustrated in Fig.~\ref{fig:linecuts}, which compares spectra recorded with and without field cycling at $\mu_0H = -25$\,mT, i.e., within the field range of the LTS. For $0^{\circ}$ and $10^{\circ}$, field cycling results in a distinct gyration mode at $f \approx 2.4$\,GHz, accompanied by a pronounced reduction of the spectral weight of the $-Q$ mode. For $20^{\circ}$, the gyration mode is still clearly resolved, whereas the changes of the $\pm Q$ modes are small. At $30^{\circ}$, no excitations characteristic of skyrmions are observed despite field cycling, consistent with the absence of signatures of the LTS in magnetometry. Accordingly, the spectral weight of the gyration mode decreases under rotation of the field away from $\langle 100\rangle$, indicating that less LTS is nucleated under field cycling, in agreement with the magnetometry data.

For $\alpha \lesssim 40^{\circ}$, the changes of slope of the Kittel mode coincide with the transition fields inferred from magnetometry, i.e., the two techniques agree well in the angular range of the tilted conical state and the LTS. For larger angles, the minimum of the Kittel mode, marking the transition between the conical and the field-polarized state, is shifted to larger absolute fields than $H_{c2}$, by roughly 15\,mT at $90^{\circ}$. As the field orientation with respect to the sample is identical in both setups, demagnetization effects cannot account for this shift. We attribute it to remanent fields and field offsets of the vector magnet used for the microwave spectroscopy, which increase with the field component perpendicular to the CPW plane.

Beyond the signatures of the LTS, two further observations are noteworthy. First, the resonances attributed to standing spin-wave modes broaden distinctly for field directions around $\alpha \approx 45^{\circ}$ and narrow again for larger angles. Second, for oblique field directions, each of the $\pm Q$ branches splits into two modes, whereas no such splitting is observed for $\alpha = 0^{\circ}$ and $90^{\circ}$. The origin of these observations is beyond the scope of this paper. As possible contributions we note that, in a cuboid sample, the internal field becomes increasingly inhomogeneous when the field is oriented obliquely with respect to the sample faces, that the character and the hybridization of the $\pm Q$ modes depend sensitively on the demagnetization factors~\cite{2015_Schwarze_NatMater}, and that magnetocrystalline anisotropies may cause a coexistence of modulations with different propagation directions for field directions away from high-symmetry axes and, as reported for field along $\langle 110\rangle$ at low temperatures, pronounced anharmonic distortions of the helimagnetic modulations~\cite{2025_Ukleev_JApplPhys}. Further, as the field is rotated with respect to the CPW, the weak out-of-plane component of the excitation field changes from transverse to longitudinal with respect to the static field, which may affect the relative spectral weights of individual modes.


\section{Discussion}

\subsection{Angular range of the tilted conical state and the LTS}

Our data establish that both the tilted conical state and the LTS are observed in a finite range of field directions around $\langle 100\rangle$, consistent with their absence for field along $\langle 110\rangle$ and $\langle 111\rangle$~\cite{2018_Chacon_NatPhys, 2018_Halder_PhysRevB}. The angular ranges, however, differ distinctly. Signatures of the tilted conical state persist up to ${\sim}40^{\circ}$, whereas signatures of the LTS are not observed beyond ${\sim}20^{\circ}$. The angular range of the tilted conical state is qualitatively consistent with neutron scattering on Zn-substituted Cu$_2$OSeO$_3$, where deviations of the propagation direction from the field emerged when the field was rotated from $\langle 111\rangle$ towards $\langle 100\rangle$~\cite{2021_Moody_PhysRevRes}. Throughout, skyrmions are observed only in combination with the tilted conical state, consistent with the latter serving as an intermediate state for the nucleation of the LTS~\cite{2018_Chacon_NatPhys, 2018_Halder_PhysRevB, 2021_Aqeel_PhysRevLett}. Our results show, in addition, that the presence of the tilted conical state is not sufficient for the formation of the LTS.

Two scenarios may account for the narrower angular range of the LTS, which cannot be distinguished by our data. On the one hand, the thermodynamic stability of the LTS may be restricted to a smaller angular range than that of the tilted conical state. The energy gain of the LTS from the cubic anisotropy depends on the orientation of the magnetization components with respect to the easy axes~\cite{2018_Chacon_NatPhys, 2018_Halder_PhysRevB} and is therefore expected to decrease when the field is rotated away from $\langle 100\rangle$. Calculations of the relative energies so far addressed the high-symmetry directions~\cite{2020_Leonov_PhysRevRes, 2023_Leonov_FrontPhys}, while, to our knowledge, calculations for arbitrary field directions are not available. On the other hand, the nucleation of the LTS may become increasingly inhibited kinetically. For field along $\langle 100\rangle$, four energetically equivalent domains of the tilted conical state form, with propagation directions tilting towards different $\langle 111\rangle$ axes~\cite{2018_Qian_SciAdv, 2018_Chacon_NatPhys}. Under rotation of the field, this degeneracy is partially lifted, which may affect the density and character of domain boundaries that putatively serve as nucleation sites~\cite{2018_Chacon_NatPhys, 2022_Leonov_PhysRevRes}.

The persistence of an LTS nucleated for field along $\langle 100\rangle$ under subsequent rotation of the crystal by up to $40^{\circ}$, as observed in neutron scattering~\cite{2018_Chacon_NatPhys}, shows that, once formed, skyrmions remain at least metastable beyond the angular range identified here. Moreover, skyrmions were observed at low temperatures for all field directions after rapid field cooling through the high-temperature skyrmion pocket~\cite{2019_Bannenberg_npjQuantumMater}, where the cooling rate required to preserve a sizable skyrmion population depends sensitively on chemical substitution~\cite{2019_Birch_PhysRevB}. Metastable skyrmion lattices at low temperatures were also created by means of electric fields~\cite{2016_Okamura_NatCommun} and studied by means of magnetic resonance~\cite{2016_Okamura_NatCommun, 2021_Takagi_PhysRevB}. The angular range reported in our study hence characterizes the formation of the LTS from the field-polarized state under field sweeps at low temperature and fixed field direction, rather than the range of field directions in which skyrmions may exist at low temperatures. Further, the signatures of the LTS are sensitive to the sample shape and hence the distribution of demagnetization fields~\cite{2018_Halder_PhysRevB}. The precise angular limits may therefore depend on the sample shape and the field history, while we expect the qualitative hierarchy of the angular ranges of tilted conical state and LTS to be robust.

The essentially unchanged field values of the signatures at $H_{s1}^{d}$ and $H_{s2}^{u}$ under rotation contrast with the pronounced shifts of the boundaries of the tilted conical state. In neutron scattering, the skyrmion plane of an existing LTS was found to remain oriented perpendicular to $\langle 100\rangle$ under rotation~\cite{2018_Chacon_NatPhys}. If the unwinding of the skyrmions were controlled by the field component along $\langle 100\rangle$, the transition fields in terms of the applied field would scale with $1/\cos\alpha$. For $H_{s1}^{d}$, the expected shift of less than 1\,mT up to $20^{\circ}$ is below the field resolution of our data. For $H_{s2}^{u}$, a shift of about 3\,mT would be expected at $15^{\circ}$, which is not observed. As the shoulder at $H_{s2}^{u}$ is weak and broad, however, this observation does not permit firm conclusions. A more detailed account of the unwinding processes under oblique fields requires microscopic information, for instance, from neutron scattering.

\subsection{Comparison of magnetometry and microwave spectroscopy}

Magnetometry and microwave spectroscopy consistently identify the same angular range of the LTS, although the nucleation was promoted in different ways. In the microwave measurements, excitations characteristic of skyrmions are observed only after cycling the field in the tilted conical state. In the magnetometry measurements, in contrast, signatures of the LTS are observed without additional cycling. This apparent discrepancy may be attributed to the AC excitation used to measure the AC susceptibility, which presumably acts as low-amplitude field cycling, akin to the small field oscillations found to increase the volume fraction of the LTS in neutron scattering~\cite{2018_Chacon_NatPhys}. At each field value, the sample is subject to several hundred cycles with an amplitude of 0.5\,mT, accumulating to a large number of cycles while the field is swept through the tilted conical state. Notably, the intensity of the resonances of the LTS was reported to increase with the number of field cycles up to several thousand cycles~\cite{2022_Lee_JPhysCondensMatter}.

The microwave spectra add information that is not accessible in magnetometry. While the signatures at $H_{s1}^{d}$ and $H_{s2}^{u}$ reflect the unwinding of skyrmions at the boundaries of the LTS, the gyration mode is observed throughout the field range in which skyrmions are present and directly confirms their skyrmionic character~\cite{2021_Aqeel_PhysRevLett}. The redistribution of spectral weight between the $\pm Q$ modes and the gyration mode thereby qualitatively reflects the relative volume fractions of the tilted conical state and the LTS. A quantitative analysis, which is beyond the scope of the present study, may be exploited in future studies of the nucleation kinetics.

\section{Conclusions}

In conclusion, we determined the dependence of the low-temperature skyrmion state and the tilted conical state in Cu$_2$OSeO$_3$ on the direction of the magnetic field by combining magnetometry and broadband microwave spectroscopy at 5\,K. Signatures of the LTS are observed only for field directions within ${\sim}20^{\circ}$ of $\langle 100\rangle$ and decrease in strength under rotation, while the field values at which the skyrmions unwind remain essentially unchanged. The tilted conical state persists up to ${\sim}40^{\circ}$, beyond which the conventional sequence of helical, conical, and field-polarized states is recovered. Hence, for the formation of the LTS under field sweeps at low temperature, the tilted conical state is a necessary but not a sufficient condition. In the microwave spectra, the counterclockwise gyration mode provides a direct indicator for the presence of skyrmions and a qualitative measure of their volume fraction, corroborating the magnetometry data. Our results motivate microscopic studies, for instance, by means of small-angle neutron scattering on spherical samples, as well as theoretical calculations for arbitrary field directions, in order to disentangle the thermodynamic stability and the kinetics of nucleation of the LTS under oblique fields.

\begin{acknowledgments}
We wish to thank H.~Berger for providing the Cu$_2$OSeO$_3$ single crystals as well as J.~Friedrich, D.~Mettus, S.~Mayr, and G.~Neumann for discussions and support. This study received funding from the Deutsche Forschungsgemeinschaft (DFG, German Research Foundation) under TRR80 (From Electronic Correlations to Functionality, Project No.~107745057), TRR360 (Constrained Quantum Matter, Project No.~492547816), SPP2137 (Skyrmionics, Project No.~403191981, Grant PF393/19), and the excellence cluster MCQST under Germany's Excellence Strategy EXC-2111 (Project No.~390814868). Financial support by the European Research Council (ERC) through Advanced Grants No.~291079 (TOPFIT) and No.~788031 (ExQuiSid) is gratefully acknowledged.
\end{acknowledgments}

\bibliography{references}

\end{document}